\documentclass[twocolumn,showpacs,prl,10pt]{revtex4}
\usepackage{amsmath,amssymb}
\usepackage{times}
\usepackage{epsfig}

\begin{document}

\title {Thermoelectric power in one-dimensional Hubbard model} 
\author{M. M. Zemlji\v c$^{1}$ and P. Prelov\v sek$^{1,2}$}
\affiliation{$^1$J.\ Stefan Institute,
SI-1000 Ljubljana, Slovenia}
\affiliation{$^2$ Faculty of Mathematics
and Physics, University of Ljubljana, SI-1000 Ljubljana, Slovenia}

\date{\today}

\begin{abstract}
The thermoelectric power $S$ is studied within the one-dimensional
Hubbard model using the linear response theory and the numerical
exact-diagonalization method for small systems. While both the
diagonal and off-diagonal dynamical correlation functions of particle
and energy current are singular within the model even at temperature
$T>0$, $S$ behaves regularly as a function of frequency $\omega$ and
$T$. Dependence on the electron density $n$ below the half-filling
reveals a change of sign of $S$ at $n_0
=0.73 \pm 0.07 $ due to strong correlations, in the whole $T$ range
considered. Approaching half-filling $S$ is hole-like and can become
large for $U \gg t$ although decreasing with $T$.

\end{abstract}  
 
\maketitle
  
\section{Introduction} 

The thermoelectric power (TEP) or the Seebeck coefficient $S$ can show
quite anomalous behavior in metals with strongly correlated
electrons. This has been recognized very clearly in the normal-state
of high-$T_c$ cuprates, where numerous experimental studies
\cite{kais,coop} reveal several unusual features of the TEP $S(T)$: a)
In the underdoped regime with small concentration of holes $n_h \ll 1$
introduced into the reference antiferromagnet (AFM) $S$ is positive
and large, showing a decrease with temperature $dS/dT<0$, except at
very low $T \sim T_c $ where the usual Fermi-liquid (FL) behavior $S
\propto T$ sets in, b) as a function of hole doping the TEP $S$ also
decreases and approaches zero or even changes sign in the overdoped
regime. The latter phenomena can be attributed to the effect of strong
correlations, since for noninteracting (or weakly interacting)
electrons one would expect negative and small $S<0$ within the same
regime.

Although one of the easiest transport quantities to measure, the TEP
proved to be a hard challenge for a theoretical consideration and
understanding. Apparently, it is even more demanding than some other
transport properties as, e.g., the electrical conductivity $\sigma$,
which are also anomalous in many correlated systems, and in
cuprates in particular.  Since the Boltzmann transport theory can fail
completely for strongly correlated electrons, the general linear
response approach has to be followed for $S$ \cite{maha}. Nontrivial
expressions have been obtained in this way in the limit of high
temperature \cite{chai} where $S(T \to \infty)$ approaches a finite
value. The limiting value of $S$ is also known for the extended
Hubbard model with an
infinite-repulsion $U \to \infty$ \cite{chai}
and for more more complicated models \cite{ktm}. An extension of
$S(T)$ to finite $T<\infty$ has been studied within the $U=\infty$
Hubbard model using the retraceable-path approximation \cite{ogur}. A
numerical analysis of $S(T)$ within the planar $t$-$J$ model has
obtained several features relevant to cuprates \cite{jprev}.
The latter calculation is based on
an observation that the ratio of off-diagonal and diagonal dynamical
correlation functions between the energy and particle current,
respectively, is nearly constant. Using the exact diagonalization of
small systems the TEP has been studied also for models with the
orbital degeneracy \cite{kosh}. Another line of approach is
to express $S(T)$ in terms of spectral functions. This is well posed
approximation within the dynamical mean-field theory \cite{pals,oudo},
valid for the Hubbard model in the limit of large dimensions. Using
spectral functions obtained via the fluctuation-exchange approximation
the latter approach has been applied also to the 2D model \cite{hild} and
further improved by the inclusion of vertex corrections \cite{kont}.

The one-dimensional (1D) Hubbard model, which is the subject of this
study, is known to possess special features relative to systems with
higher $D>1$.  As 1D models of interacting electrons in general, it
complies to the phenomenology of Luttinger liquids and exhibits the
spin-charge separation \cite{schu}. Using the Bethe-Ansatz solution
for low-energy spinon and holon excitation branches, the low-$T$
behavior of $S$ has been considered in Ref.\cite{staf}. Results appear
quite unusual since $S$ is claimed to remain positive at all electron
densities $n<1$ below half filling.
This study indicates that even $S(T \to 0)$ is an
open question within the 1D Hubbard model, and even more the behavior at
higher $T$. On the other hand, it has been recognized that the
integrability of the 1D Hubbard model implies singular transport
quantities \cite{cast,zoto,zprev}. In particular, it has been shown
that away from half-filling, i.e., for $n \neq 1$, the charge and
energy (heat) transport remain dissipationless even at $T>0$, i.e., in
the d.c. limit the electrical as well as the thermal conductivity both
diverge, $\sigma_0 \to \infty$ and $\lambda_0 \to \infty$
\cite{zoto}. Since $S(T)$ is the ratio of two correlation functions,
it is expected to remain well defined and finite
\cite{zprev}.

In order to clarify the behavior of the TEP within the 1D Hubbard
model, we present in the following results of the numerical study of
the dynamical $\tilde S(\omega,T)$, based on the linear response
theory, generalized to allow for singular transport quantities.
Dynamical response functions are evaluated numerically for finite
chains with $L$ sites using the approach of exact diagonalization.  In
particular, for the largest systems we apply the finite-temperature
Lanczos method (FTLM) \cite{jprev} which enables us to consider a model
with up to $L=14$ sites . We concentrate on the regime of strong
correlations $U \geq 2~t$.

The paper is organized as follows: In Sec.~2 we present the general
linear response formalism for transport quantities within the 1D
Hubbard model, with the special care for singular behavior in
the d.c. limit $\omega \to 0$. Different limits are also
considered. Results of the numerical study of finite systems are
presented in Sec.~3 and conclusions are given in Sec.~4.

\section{Linear response theory}

We  consider the 1D Hubbard model 
\begin{equation}
H=-t \sum_{l,s} (c_{l+1,s}^{\dagger} c_{l, s} + \mathrm{H.c.})  +
U \sum_l n_{l\uparrow} n_{l\downarrow} , \label{eqh}
\end{equation} 
with the nearest neighbour (n.n.) hopping term $H_t$ and the onsite
repulsion $U$ on a chain of $L$ sites. The particle current has the
usual form (we set further $\hbar=1$ and the lattice spacing $a_0=1$),
\begin{equation}
J_n=t \sum_{l,s} (i c_{l+1,s}^{\dagger} c_{ls} + \mathrm{H.c.}), 
\label{eqjn}
\end{equation}
whereas the energy current can be expressed in terms of n.n. energy
operators \cite{beni,zoto}
\begin{equation}
J_E=i \sum_l [h_{l},h_{l+1}], \qquad H=\sum_l h_{l}. \label{eqje1}
\end{equation}
It then follows \cite{zoto} from Eq.~(\ref{eqh})
\begin{eqnarray}
&&J_E=-t^2 \sum_{l,s} (i c_{l+1,s}^{\dagger} c_{l-1,s} + \mathrm{H.c.})
+ \nonumber \\ &&+\frac{Ut}{2} \sum_{l,s}\{i(c_{l+1,s}^{\dagger}
c_{ls}+ c_{ls}^{\dagger} c_{l-1, s})n_{l,-s} + \mathrm{H.c.}\}.
\label{eqje2}
\end{eqnarray}

Within the linear response formalism we calculate general transport
coefficients $L_{ij}$ \cite{maha}, defined by $J_i = \sum_j L_{ij}
F_j$. Here, $i,j=n,Q$, where $J_Q=J_E-\mu J_n$ is the heat current,
$\mu$ is the chemical potential, and related generalized forces are
given by $F_n=-(\nabla \mu + e_0{\cal E})/T, F_Q = -\nabla T/T^2$.
Since the transport coefficients in the 1D Hubbard model are singular in
the d.c. limit \cite{zoto,zprev}, it is essential to generalize the
formalism to finite frequencies, dealing with response functions
$L_{ij}(\omega)$. If one introduces appropriate polarization operators
\cite{mald,loui} $P_n=\sum_l l n_l$, $P_E=\sum_l l h_l$ and
$P_Q=P_E-\mu P_n$ (valid for open boundary conditions), we can write
within the linear response theory
\begin{equation}
L_{ij}(\omega)=\frac{T}{i \omega^+} ( \langle \tau_{ij} \rangle - 
\chi_{ij}(\omega) ), \label{eql}
\end{equation}
where $\omega^+=\omega+i\epsilon$,
\begin{equation}
\tau_{ij}=-\frac{i}{L} [J_i,P_j], \label{eqtau}
\end{equation}
 and
\begin{equation}
\chi_{ij}(\omega)=-\frac{i}{L}\int_0^\infty dt e^{i\omega^+ t}
\left<[J_i(t),J_j(0)]\right>. \label{eqchi}
\end{equation}
It follows from Eq.(\ref{eqtau}) that $\tau_{nn}=-H_t/L$ \cite{mald},
and
\begin{eqnarray}
&& \tau_{nE}=\frac{2 t^2}{L} \sum_{l,s} (c_{l+1,s}^{\dagger} c_{l-1,s} +
\mathrm{H.c.})  - \nonumber \\ &&-\frac{Ut}{2 L}
\sum_{l,s}\{(c_{l+1,s}^{\dagger} c_{ls}+ c_{ls}^{\dagger} c_{l-1,
s})n_{l,-s} + \mathrm{H.c.} \},
\label{eqtaune}
\end{eqnarray}
while $\tau_{nQ}=\tau_{nE}-\mu \tau_{nn}$. Note that the dynamical
electrical conductivity, more frequently discussed in
this context, is given by $\sigma(\omega)=e^2_0 L_{nn} (\omega)/T$.

Finally, we get for the dynamical TEP \cite{maha}
\begin{eqnarray}
\tilde S(\omega)&=&-\frac{1}{e_0T} \frac{L_{nQ}(\omega)}
{L_{nn}(\omega)} = \nonumber \\ &=&
-\frac{1}{e_0T} \Bigl( \frac{\langle \tau_{nE}\rangle-
\chi_{nE}(\omega)} {\langle \tau_{nn}\rangle - \chi_{nn}(\omega)}-\mu
\Bigr).
\label{eqsom}
\end{eqnarray}

\noindent In most model systems (and in real materials) we are dealing with the
situation where transport coefficients at $T>0$ are not singular
within the d.c. limit, i.e., $L_{ij}(\omega \to 0)=L^0_{ij}$ are
finite.  This is related to the ergodic behavior of the system and
Eq.(\ref{eql}) in this case requires the equality of the d.c. adiabatic and
isothermal susceptibilities, $\chi_{ij}(\omega\to 0)=\chi_{ij}^T\equiv
\langle \tau_{ij}\rangle$ \cite{loui}.

It has been shown that in 
this respect the 1D Hubbard model is singular \cite{zoto}, as well as
some other 1D integrable quantum many-body models \cite{zprev}. Let us
define the generalized stiffness
\begin{equation}
D_{ij}= \frac{1}{2} (\langle\tau_{ij}\rangle- \chi_{ij}(\omega \to 0)).
\label{eqd}
\end{equation}
Within the 1D Hubbard model $D_{ij}(T>0)$ remain
finite away from half-filling $n \neq 1$ \cite{zoto}. The general
argument has been given with the finite overlap of current operators
$\langle J_i Q_n \rangle$ where $Q_n, n=1,L$ are conserved quantities
due to integrability. Note that in a macroscopic system there is an
infinite number of operators $Q_n$. In particular, $Q_3$ corresponds
closely to $J_E$, obtained by the replacement in Eq.(\ref{eqje2}) $U/2 \to
U$. In spite of singular $L_{ij}(\omega \to 0)$ the d.c. TEP $S=\tilde
S(\omega\to 0)$ is expected to remain well defined and finite.
From the representation chosen in
Eq.(\ref{eqd}) we can express $S$ as
\begin{equation}
S = - \frac{1}{e_0T} \Bigl(\frac{D_{nE}}{D_{nn}} -\mu \Bigr).
\label{eqs0}
\end{equation}

There are some limits which give a nontrivial test to our results
presented below. For {\it noninteracting electrons} at $T>0$ it is
straightforward to evaluate $S$. Since $J_n$ and $J_E$ are conserved
quantities (constants of motion), we get from Eq.(\ref{eqchi}) that
$\chi_{nn}(\omega)=\chi_{nE}(\omega)=0$ and
\begin{eqnarray} 
\tau_{nn}&=&\int f(\epsilon)g(\epsilon) \epsilon d\epsilon, \nonumber \\
\tau_{nE}&=&\int f(\epsilon)g(\epsilon) \left(2 \epsilon^2-4t^2\right)
d\epsilon, \label{eqtauf}
\end{eqnarray}
where $g(\epsilon)=2 (4t^2-\epsilon^2)^{-1/2}/\pi $,
$f=1/(\exp[(\epsilon-\mu)/T]+1)$ and $\mu$ is fixed by the density
$n=\int f(\epsilon) g(\epsilon) d\epsilon$.  

In {\it the limit of large $k_B T\gg (U,t)$} the result coincides with
that of noninteracting electrons at $k_B T \gg t$,
\begin{equation}
S=-\frac{k_B}{e_0}\ln{\frac{2-n}{n}}. \label{eqsf}
\end{equation}

On the other hand, a nontrivial generalized Heikes formula is obtained for the
case of {\it large $(k_B T,U)\gg t$ but $k_BT \ll U$} \cite{chai},
\begin{equation}
S=-\frac{k_B}{e_0}\ln{\frac{2(1-n)}{n}}. \label{eqsh}
\end{equation}

It has been recently observed \cite{jprev} that in the most frustrated
regime of strongly correlated systems the transport coefficients
$L_{nE}$ and $L_{nn}$ are closely related. Numerical results for the
planar $t$-$J$ model thus reveal (at the intermediate doping of the
reference antiferromagnet) that in a broad range of $\omega,T$ the ratio
$R(\omega)=L_{nE}(\omega)/L_{nn}(\omega)$ is nearly constant.
Extending this behavior to $T \to 0$ a simple relation for $S(T)$
would follow, which is very attractive for application in
complicated correlated systems \cite{tohy}. It is of interest to which
extent within the 1D Hubbard model $R(T)=D_{nE}/D_{nn}$ behaves similarly.

\section{Results} 

We analyse dynamical transport coefficients $L_{nn}(\omega)$ and
$L_{nE}(\omega)$ and consequently $\tilde S(\omega)$ by performing the
exact diagonalization of the model Eq.~(\ref{eqh}) on a finite chain
with $L$ sites and with periodic boundary conditions.  Taking into
account the number of electrons $N_e$ and the total spin projection $S^z$ as
conserved quantities as well as the translation symmetry (wavevector
$q$) systems up to $L =10$ sites (all fillings) can be fully
diagonalized to get results at arbitrary $T$. Using the FTLM one can
analyse larger systems at $T>0$ \cite{jprev} and in the following we
present results obtained for $L=12, 14$ with the number of basis
states within a symmetry sector up to $N_{st} \sim 800.000$.
All $N_e \le L$ (for $N_e>L$
we employ the electron-hole symmetry of the model) are investigated in
order to perform the grand-canonical averaging which allows for a
continuous variation of $n=N_e/L$. In the present FTLM application we
use typically $M=100$ Lanczos steps and random averaging over $R \sim
20$ samples within each symmetry sector.

Within FTLM, as well as in any exact calculation of a small system, it
is important to realize that results have a restricted thermodynamic
validity due to finite-size effects, which begin to dominate at low
$T<T_{fs}$. Clearly, $T_{fs}$ depends on the size of the system, but
as well on the density and character of low-$T$ excitations. It is
characteristic that $k_BT_{fs}/t
\sim 0.2$ (at $n \sim 1$) in the 1D Hubbard chain with $L=14$ is larger
that in a corresponding 2D (square lattice) Hubbard model with
$N=4\times 4=16$ sites \cite{bonc} (where it was found $k_BT_{fs}/t \sim
0.1$ at the intermediate doping of the reference insulator) in spite of an
evidently shorter linear extension. The difference can be explained
with the anomalous behavior of the planar Hubbard model (in the
strong-correlation regime $U \gg t$) \cite{bonc} and of related 2D
$t$-$J$ model \cite{jprev} which show a non-FL behavior close to
half-filling, i.e., a large degeneracy of low-lying states results
in a large entropy and small $T_{fs}$. On the other hand, the 1D
Hubbard model at low $T$ follows the phenomenology of Luttinger
liquids \cite{schu}, with well defined excitation branches of spinons
and holons, resulting in an increased but not anomalous entropy.

\vskip 1.0truecm
\begin{figure}[htb]  
\centering \epsfig{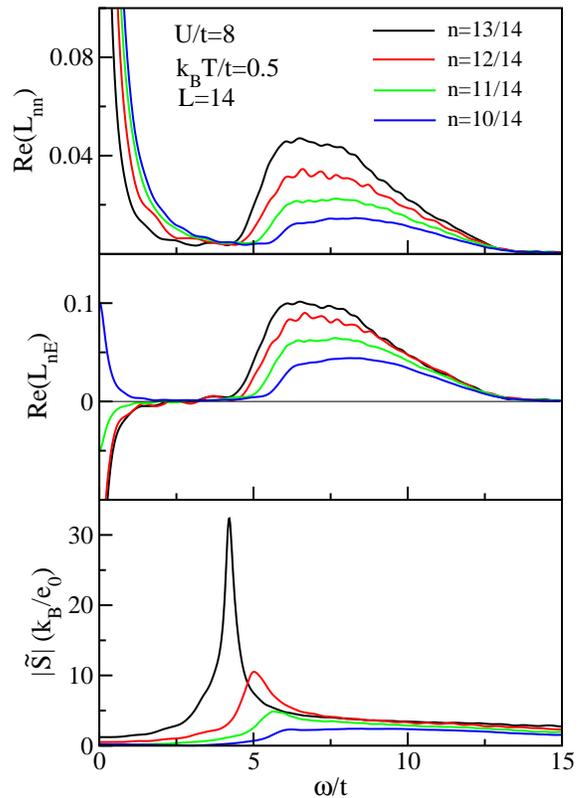}
\caption{Transport coefficients Re~$L_{nn}(\omega)$,
Re~$L_{nE}(\omega)$ and TEP $|\tilde S(\omega)|$, calculated within
the 1D Hubbard model with $U/t=8$, $L=14$, $T=0.5~t$, for different
electron densities $n$ close to half-filling. Results obtained via
FTLM are broadened with $\delta=0.25~t$.}
\label{fig1} 
\end{figure}

Let us first discuss the frequency-dependent $L_{nn}$, $L_{nE}$ and
$\tilde S$. In Fig.~1 we present Re~$L_{nn}(\omega) \propto \sigma$,
Re~$L_{nE}(\omega)$ as well as $|\tilde S(\omega)|$, calculated using the
FTLM in a $L=14$ system for the typical strong-correlation regime
$U=8~t$ for fixed $k_BT=0.5~t$ and different $n$ close to half-filling.
Spectra are smoothed with an additional broadening $\delta=0.25~t$. As already
found in previous studies \cite{zoto,zprev}, finite $D_{nn}(T>0)$
(away from half-filling $n \ne 1$) is reflected in a singular
(delta-function) contribution to Re~$L_{nn}(\omega \to 0)$ in Fig.~1. The
incoherent part pronounced for $\omega>4t$, mainly due to the
transition from the lower to upper Hubbard band, decreases away from
$n \sim 1$, consistent with the free electron propagation when approaching
the empty band, $n \to 0$. An analogous structure is observed in the
off-diagonal $L_{nE}(\omega)$, whereby the singular contribution
$D_{nE}$ can be negative as well, and in fact shows the change of sign
upon decreasing $n$. The resulting (absolute value) $|\tilde
S(\omega)|$ evaluated from Eq.(\ref{eqsom}) is presented in Fig.~1.
In the strong-correlation regime it seems to be characteristic (close to
half-filling) that $\tilde S(\omega)$ shows a pronounced resonance at
$\omega \sim \omega_0$ with $\omega_0$ linked to the pseudo-gap in the
incoherent part. The resonance is determined by the minimum of
$|L_{nn}(\omega_0)|$, i.e., by the vanishing Im~$L_{nn}(\omega_0)=0$
whereby also Re~$L_{nn}(\omega_0)$ is small.  This results in a sharp
peak in $|\tilde S(\omega)|$, with a simultaneous change of phase in
complex $\tilde S(\omega)$. The same effect we find for $U=4~t$. Away
from $\omega \sim \omega_0$ the dynamical TEP $|\tilde S(\omega)|$ behaves
quite smoothly approaching both (different) nontrivial limits
$\omega=0$ and $\omega \to \infty$.

\vskip 0.5truecm
\begin{figure}[htb]
\centering
\epsfig{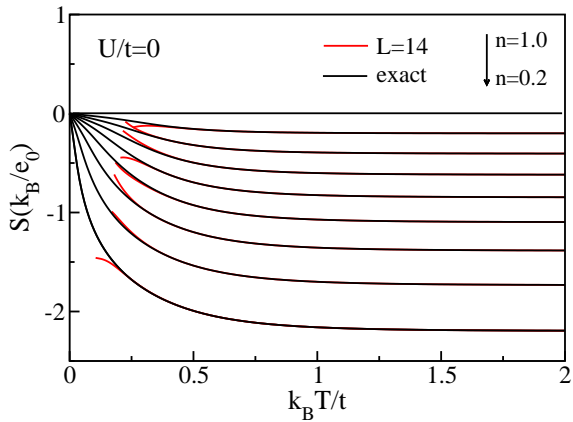}
\caption{TEP $S$ (in units $k_B/e_0$) vs. $T$ for
noninteracting electrons for different densities $n$, calculated from the
analytical result (\ref{eqtauf},\ref{eqs0}), compared
to the result for a small system $L=14$.}
\label{fig2}
\end{figure}

In Fig.~2 we present the reference d.c. TEP $S(T)$ for noninteracting
fermions, evaluated from Eq.~(\ref{eqtauf},\ref{eqs0}) for different
$n$. We also display the result of a grand-canonical calculation of
$S$ on a finite-size system with $L=14$ sites. The deviations of the
latter from exact results at low $T$ can serve as an estimate for
$T_{fs}$, relevant for FTLM calculations presented below.  It should
be noted that in general at $T \to 0$ small-system results for $S$ do
not get a meaningful limit, in particular they do not reproduce
correct FL $S \propto T$ behavior. It follows from Fig.~2 that
$k_BT_{fs} \sim 0.2-0.3~t$, depending on $n$. Note also that $S \equiv 0$
for $n=1$ due to the particle-hole symmetry of the model, which
remains valid also for general $U \ne 0$ \cite{ben2}. It should also
be observed from Fig.~2 that the FL regime with $S \propto T$ is
restricted to quite low $k_BT<0.4~t$ even for noninteracting fermions, in
spite of much broader band $W=4~t$.

\begin{figure}[htb]
\centering
\epsfig{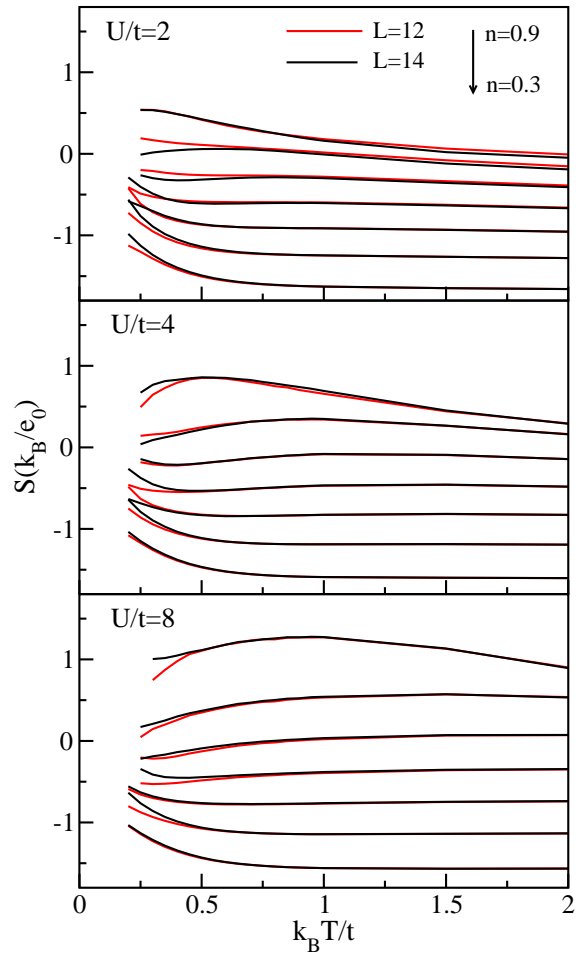}
\caption{$S$ vs. $T$ for the 1D Hubbard model with $U/t=2,4,8$,
respectively, as calculated for different $n$ with the FTLM for $L=12, 14$
chains.}
\label{fig3}
\end{figure}

FTLM results for $S(T)$ for different $U/t=2,4,8$ and $n=0.3 - 0.9$
are presented in Fig.~3. For a given system of $L$ sites and fixed $T$
we calculate using FTLM $S(\omega \to 0)$ values for all sectors
$1<N_e<L$ (taking into account the symmetry of the model for $L+1<N_e<2L-1$) and
perform grand-canonical averaging to get a continuous variation
$S(T,n)$. We show in Fig.~3 results for two systems,
$L=12, 14$, which allow to estimate $T_{fs}$ for $U>0$ cases.

Several observations can be made on the basis of $S(T)$ results in
Fig.~3: a) For $n \leq 0.6$ we get for all considered $U/t$ $S(T)<0$,
with values essentially identical to that of noninteracting electrons
in Fig.~2. b) The effect of $U>0$ becomes well pronounced close to
half-filling, where the TEP becomes hole-like, i.e. $S>0$, and at the
same time large. At large $U \gg t$, the TEP $S(T)$ stays constant in
a large $T$ window, whereas for smaller $U$, e.g. $U/t=2$, it is
suppressed already for modest $k_B T \sim t$ to the noninteracting value $S
\sim 0$. c) It is rather difficult to reach the FL regime where $S
\propto T$. Still, away from half-filling, i.e. for $n<0.6$, numerical
results appear consistent with the behavior $S(T \to 0) \to 0$. This
is, however, not evident for $n \sim 1$, in particular not for largest
$U=8~t$.

\begin{figure}[htb]
\centering
\epsfig{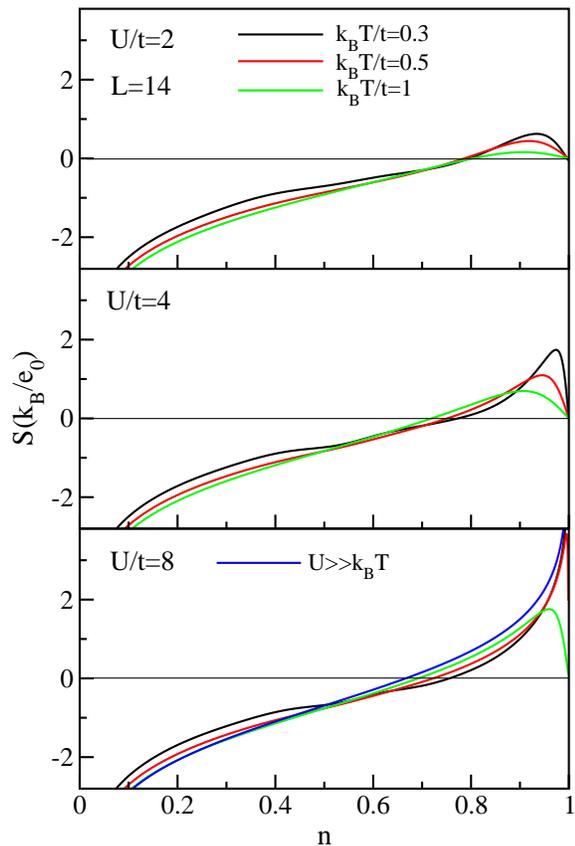}
\caption{$S$ vs. $n$ for $U/t=2,4,8$ and different $T$, as
calculated with FTLM for $L=14$ chains. For $U/t=8$ case we present
also the Heikes value, Eq.(\ref{eqsh}).}
\label{fig4}  
\end{figure}

In Fig.~4 we display the alternative representation of the TEP results
$S(n)$ (for $L=14$ system only) for various $k_B T/t=0.3, 0.5, 1.0$
and same $U/t$ as in Fig.~3.  Again, for $n<0.6$ the TEP is
electron-like $S<0$ and behaves as for noninteracting electrons. The
interesting part is for $n>0.6$ where $U>0$ induces a transition to a
hole-like $S>0$ for $n>n_0$.  In general, $n_0$ is dependent both on
$T$ and $U$. The limiting value, reached for $U \gg k_B T \gg t$, is
given by the generalized Heikes formula, Eq.(\ref{eqsh}), with the value
$n_0=2/3$. Moreover, as seen in Fig.~4, $n_0$ remains nearly constant,
i.e. $n_0\sim 0.7$, in a broad range of $T$ in the strong-correlation
regime $U/t \geq 4$. Only for a weak repulsion $U/t=2$ case the
crossing starts to approach the noninteracting $n_0 =1$. Since at
$n=1$ we get $S=0$ for all $T$, there must be a maximum $S^*=S(n=n^*)$
within the hole-like regime. The location $n^*$ as well as the value
$S^*$ depends strongly on $U$ and $T$. The limit,
Eq.(\ref{eqsh}), gives $n^*=1$ with divergent $S^* \to
\infty$. For $U=8~t$ our results  for $k_B T\leq 0.5~t$ show a good 
overall agreement with the latter limiting behavior. Note that here
due to large charge gap at $U=8~t$, the maximum appears very close to
half-filling, i.e.,  $n^* \to 1$ with $S^* \gg k_B/e_0$. With decreasing
$U/t$ as well as with increasing $T$ the maximum becomes less
pronounced and moves away from $n^*=1$.

\section{Conclusions}

Above we have presented results for the dynamical and d.c. TEP within
the 1D Hubbard model, as calculated by using the FTLM for small
systems with up to $L=14$ sites. Several comments and conclusions are
in order:

\noindent a) The main restriction in the validity of results comes
from the use of small systems. Finite-size effects in $S$ and other
quantities start to dominate results for $T<T_{fs}$. Although the FTLM
algorithm involves also the random sampling over initial
configurations \cite{jprev}, it does not essentially increase the
numerical error of $S$ and only slightly increases $T_{fs}$.

\noindent b) It is quite characteristic that observed $T_{fs}$ is
weakly dependent on $U$ and $n$, i.e., $k_BT_{fs}/t \sim 0.2-0.3$ for
$L=14$. This is strikingly different from the Hubbard model (or
$t$-$J$ model) in 2D, where $T_{fs}$ varies substantially with
concentration $n$ in the strong-correlation regime \cite{bonc,jprev},
i.e., $T_{fs}$ has a minimum at the 'optimum' doping where the quantum
antiferromagnet is frustrated by mobile holes. The latter regime is
characterized by the large entropy, hence also low $T_{fs}$. In the 1D
Hubbard model there seems to be no such phenomenon, consistent with
the low density of excitations (spinons and holons) within the
Luttinger liquid.

\noindent c) We have investigated only the regime below 
half-filling $n<1$, since due to the particle-hole symmetry of the
model, Eq.~(\ref{eqh}), it follows $S(2-n)=-S(n)$.

\noindent d) Independent of $U$ at low filling $n<0.6$ the TEP is 
electron-like, $S<0$, and essentially equal to that of noninteracting
electrons in the whole regime of $T$ considered. Although the result
appears plausible, it contradicts the conclusion based on the exact
Bethe-Ansatz solutions \cite{staf} that for large $U>t$ the TEP is
hole-like $S>0$ for all $n<1$.

\noindent e) TEP becomes strongly dependent on $U$ as well as on $T$
when approaching the half-filling. In particular, it changes the sign at
$n_0$, which in a broad regime of $U,T$ stays close to the Heikes
result $n_0 \sim 2/3$, but increases towards $n_0 \to 1$ with the
vanishing $U \to 0$. $S(n)$ reaches a maximum at $n=n^*$, located in
the regime $n_0<n^*<1$. Both $n^*$ and $S^*$ are dependent on $U$ and
$T$. In particular, $S^*$ can become very large in the
strong-correlation regime $U \gg t$ where $n^* \to 1$ and $S(n<n^*)$
follows Eq.~(\ref{eqsh}) down to low $T$.

\noindent f) The low-$T$ variation of the TEP is expected to follow
the standard FL behavior $S \propto T$. It appears quite difficult to
enter and investigate the latter regime using small-system
diagonalization. The linear dependence is restricted to a rather
narrow window $k_BT<0.5~t$ even for free fermions, as seen in Fig.~2. For
$U \gg t$ close to the half-filling this window is expected to become
even narrower, since the relevant energy scale is that of spinons with
the characteristic exchange coupling $J=4 t^2/U$. So it is not
surprising that our results in Fig.~3 for largest $U=8~t$ at $n \sim
1$ do not reveal the onset of the linear regime, down to $k_BT_{fs}\sim
0.3~t $. In any case, the low-$T$ behavior of the TEP remains an open
question which could be possibly settled by an analytical calculation
\cite{staf}.

\noindent g) It is a relevant question to what extent the
behavior within 1D Hubbard model reflects the general features of
transport and of TEP in higher dimensional systems with strongly
correlated electrons. Of particular interest is the relation with the
physics of quasi-2D high-$T_c$ cuprates and with their anomalous
electronic properties \cite{kais,coop}, which have been the motivation
of the intensive research of microscopic models of correlated
electrons. Apart from results based on the approximation of infinite
dimensions \cite{pals,oudo}, there are few model calculations of $S$
in 2D systems \cite{jprev} and the question of the variation of $S(T)$
in prototype models is not settled.  Still, some differences and
similarities between 1D and higher D systems seem to be evident.
Transport in 1D Hubbard model is specific due to its integrability,
showing in the divergence of d.c. coefficients $L_{ij}(\omega \to
0)$. Nevertheless, the TEP $S$ is well behaved and even some deviation
from integrability (e.g., by adding additional next n.n. hopping
in $H$) should not influence its behavior dramatically. The difference
between 1D and 2D (or higher D) models is in the nature of low-lying
excitations. While in the 1D model the latter follow the scenario of
the Luttinger liquid, the situation in 2D is more involved and not
fully understood yet. At least numerical investigation reveal a very
degenerate quantum state close to optimum doping
\cite{jprev,bonc}. One consequence appears to be nearly $\omega$ and
$T$-independent ratio $R=L_{nE}/L_{nn}$ \cite{jprev}. In contrast, in
1D model $R=D_{nE}/D_{nn}$ remains rather $T$-dependent (roughly as
much as $\mu$), so $S$ cannot be well expressed solely in terms of
$\mu(T)$.

\noindent h) Nevertheless, there are important similarities between
the TEP in 1D Hubbard model and available 2D results \cite{jprev} and
moreover experimental results within the normal state of cuprates
\cite{kais,coop}. In the strong correlation regime $U\gg t$, $S$
becomes hole-like for $n\le 1$. In a broad range of $T$ the change of
sign appears at $n_0 = 0.73 \pm 0.07$, close to the limiting Heikes
value $n=2/3$. At the same time the maximum value $S^*>0$ can become
large and remains as such down to quite low $T$, giving the TEP in
strongly correlated system also the potential applicability. At the
same time, in the hole-like regime the $T$ dependence of the TEP is
quite pronounced. In particular, $dS/dT<0$ at $T>T^*$, also consistent
with experiments on cuprates, although $T^*$ appears higher than in 2D
systems \cite{jprev}.
\newline

Authors acknowledge the support of the Ministry of Education, Science
and Sport of Slovenia under grant Pl-0044.

 \end{document}